\documentclass[10pt]{article}
\usepackage{amsmath}
\usepackage{bm}
\usepackage{graphicx}

\begin{document}

\title{Fast construction of the Kohn--Sham response function for molecules}




\author{%
 Peter Koval\textsuperscript{\textsf{\bfseries 1}},
 Dietrich Foerster\textsuperscript{\textsf{\bfseries 2}}
 Olivier Coulaud\textsuperscript{\textsf{\bfseries 3}}
}

\maketitle



\noindent
 \textsuperscript{1}\,CNRS, HiePACS project, INRIA Sud-Ouest, 351 Cours de la Liberation, 33405, Talence, France \\
 \textsuperscript{2}\,CPMOH, Universite de Bordeaux 1, 351 Cours de la Liberation, 33405, Talence, France \\
 \textsuperscript{3}\,INRIA Bordeaux Sud-Ouest, HiePACS project, 351 Cours de la Liberation, 33405, Talence, France



\abstract{The use of the LCAO (Linear Combination of Atomic Orbitals) method
for excited states involves products of orbitals that are known to be
linearly dependent. We identify a basis in the space of orbital products
that is local for orbitals of finite support and with a residual error that
vanishes exponentially with its dimension. As an application of our
previously reported technique we compute the Kohn--Sham density response
function $\chi_{0}$ for a molecule consisting of $N$ atoms in
$N^{2}N_{\omega }$ operations, with $N_{\omega}$ the number of frequency
points. We test our construction of $\chi_{0}$ by computing molecular
spectra directly from the equations of Petersilka--Gossmann--Gross in
$N^{2}N_{\omega }$ operations rather than from Casida's equations which
takes $N^{3}$ operations. We consider the good agreement with previously
calculated molecular spectra as a validation of our construction of
$\chi_{0}$. Ongoing work indicates that our method is well suited for the
computation of the GW self-energy $\Sigma=\mathrm{i}GW$ and we expect it
to be useful in the analysis of exitonic effects in molecules.}

\bigskip

Accepted for publication in \textbf{Physica Status Solidi}, 05.02.2009 both
as contribution to the Proceedings of TNT2009 conference (Barcelona, Spain) and
as a featured article.

\section{Introduction}

The method of ``linear combination of atomic orbitals'' (LCAO) goes back to
the early days of quantum mechanics \cite{Mulliken:1967} and remains a good
choice in ab-initio calculations of molecules or solids. LCAO provides a
parsimonious basis for representing molecular orbitals and one-particle
Green's functions. However, electronic structure theory also involves the
electronic density and the expression for the electronic density contains
all non vanishing products of orbitals, a set of quantities that are known
to be linearly dependent.

The many existing constructions of a basis in the space of products may be
divided into two classes. A first type of construction, called
``resolution of identity'' method was
proposed by Boys and Shavitt \cite{Boys-Shavitt:1959} and focuses on the
representation of the electronic interaction, see also Casida \cite{Casida}.
For Slater-type functions a \textquotedblleft density
fitting\textquotedblright\ procedure was developed by Baerends
\textit{et al} \cite{Baerends-etal:1973,Te-Velde-etal:2001}. The goal
of both of these methods is to represent the electronic density
(a two-center quantity within LCAO method) in a basis of one-center functions.
For a good discussion of these techniques, see \cite{Skylaris-etal:2000}.

A second class of methods provides a basis for both response functions and
interactions. For solids and within the muffin-tin approach,
a ``product basis'' was proposed by
Aryasetiawan and Gunnarsson \cite{Aryasetiawan-Gunnarsson:1994} who removed
the linear dependence by orthogonalizing an overlap matrix.
Gaussian-type auxiliary functions were also used in solid state theory \cite%
{Rohlfing-etal:1995}. Blase and Ordej\'{o}n \cite{Blase-Ordejon:2004} used
Gram-Schmidt orthogonalization to eliminate the linear dependence from
products on the same atom.

Our own construction \cite{DF:2008,DF:2009} belongs to the second class of methods.
It was developed in the context of LCAO for orbitals of finite support and
keeps the locality properties of the underlying atomic orbitals.
In the present paper we further test our calculational framework by
computing spectra of medium sized molecules and we also give an expression
of the GW self-energy in our product basis.

\begin{figure*}[ht]
\centerline{
\includegraphics[width=7.5cm,viewport=70 60 400 300]{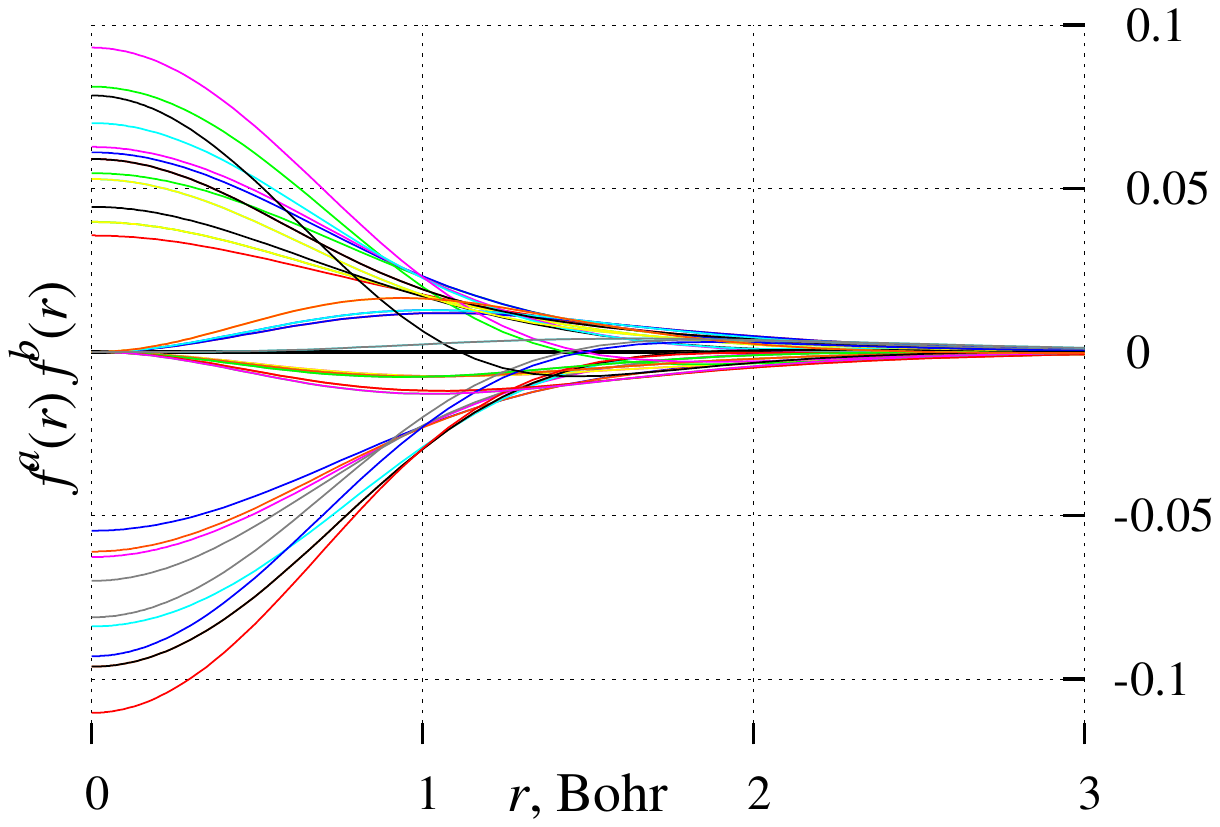}\hspace{1cm}
\includegraphics[width=7.5cm,viewport=70 60 400 300]{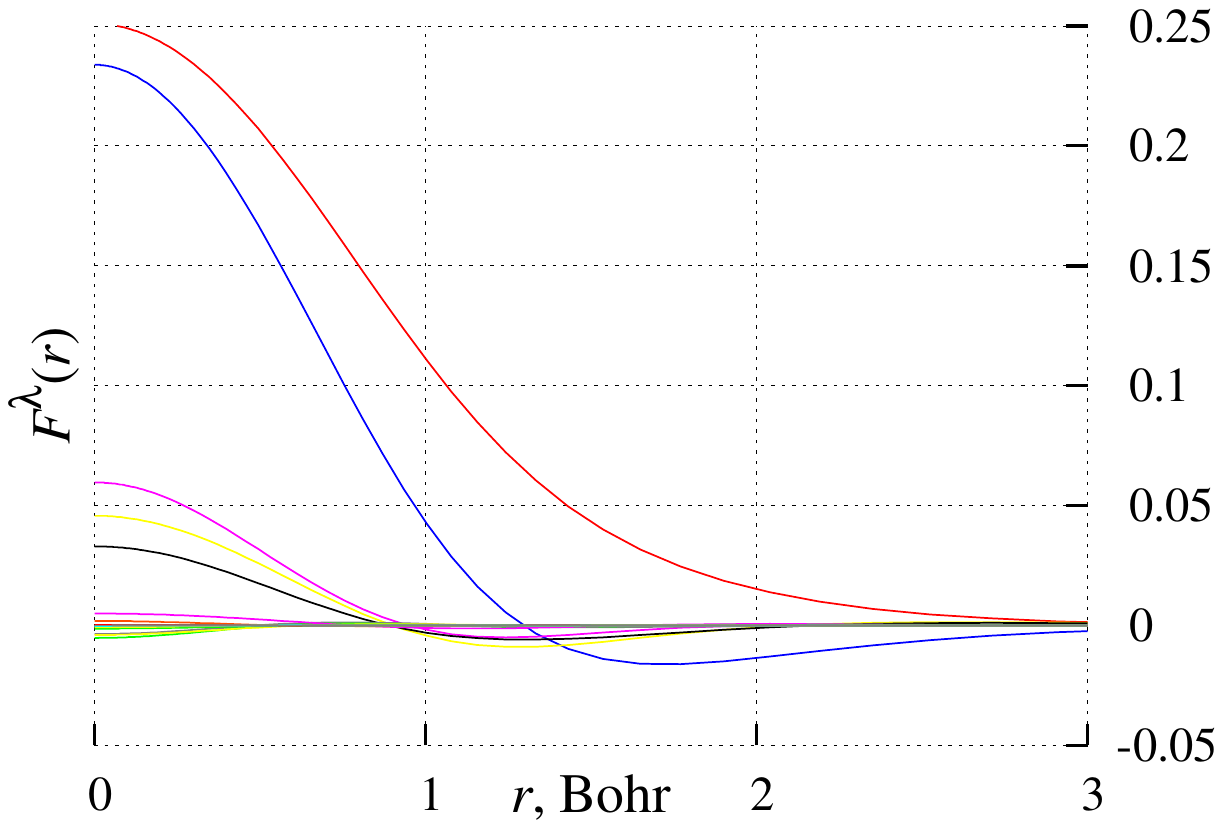}}
\caption{On the left panel, the orbital products (of total azimuthal angular momentum $m_z=0$)
are shown along the line connecting two (carbon) atoms. On the right panel, the corresponding
dominant products are shown. The origin of x-axis is the middle point between atoms (on both
panels). The linear dependence of the original orbital products and the small size of
the dominant products basis are clearly visible.}
\label{f:cc-prod}
\end{figure*}

This paper is organized as follows. In the next section, we describe our
construction of a basis in the space of products in terms of ``dominant
products''. A construction of the Kohn--Sham response function is described
in section \ref{s:response}, and the application of this response function
to the computation of molecular spectra is given in section \ref{s:tddft}.
Finally, in section \ref{s:gw}, we derive an expressions for the self-energy
in Hedin's GW approximation in our basis and give our conclusions.

\section{Reducing the number of orbital products}

\label{s:reducing}

Restated in the succinct language of second quantization, the LCAO method
consists of an expansion of the electron creation and annihilation operators 
$\psi ^{+}(\bm{r},t)$, $\psi (\bm{r},t)$ in terms of electron operators $%
c_{a}(t)$ that belong to atomic orbitals 
\begin{equation}
\psi ^{+}(\bm{r},t)\sim \sum_{a}f^{a}(\bm{r})c_{a}^{+}(t).
\label{annihilation-operator}
\end{equation}%
The functions $\{f^{a}(\bm{r})\}$ are atom centered orbitals, the operator $%
c_{a}^{+}(t)$ create localized electrons in these orbitals and the symbol $%
\sim $ alludes to the question of completeness of these LCAO orbitals, a
difficulty we shall ignore in the following. Most molecular response
functions involve the electronic density $n(\bm{r},t)$ and, therefore, the
square of the previous expansion

\begin{equation*}
n(\bm{r},t)=\psi ^{+}(\bm{r},t)\psi (\bm{r},t)\sim \sum_{a,b}f^{a}(\bm{r}%
)f^{b}(\bm{r})c_{a}^{+}(t)c_{b}(t).
\end{equation*}%
It is well known in quantum chemistry that the products of orbitals $\{f^{a}(%
\bm{r})f^{b}(\bm{r})\}$ that parametrize this density are linearly
dependent. A good illustration of this fact is provided \cite{Harriman:1986}
by the harmonic oscillator and its Hermite wave functions $\phi
_{n}(x)=H_{n}(x)\mathrm{e}^{-x^{2}/2}$ where $\frac{N(N+1)}{2}$ products $%
\{\phi _{m}(x)\phi _{n}(x)\}_{m,n=1..N}$ span a space of only $2N+1$
dimensions. The set of products $\{f^{a}(\bm{r})f^{b}(\bm{r})\}$ is usually
parametrized by extra auxiliary fitting functions 
\cite{Baerends-etal:1973,Te-Velde-etal:2001,Fitting_Functions}. %
In this paper, we use an alternative procedure that involves no fitting
functions whatsoever \cite{DF:2008}. We proceed in two steps:

\begin{itemize}
\item for each pair of atoms, the orbitals of which overlap, we enumerate
all products $F^{M}(\bm{r})=f^{a}(\bm{r})f^{b}(\bm{r})$,

\smallskip

\item we compute their metric or matrix of overlaps $G^{MN}$, diagonalize
this matrix and employ its eigenvectors and eigenvalues to define \textit{%
dominant products} $F^{\lambda }(\bm{r})$ 
\begin{eqnarray}
G^{MN} &=&\int \frac{F^{M}(\bm{r})F^{N}(\bm{r}^{\prime })}{|\bm{r}-\bm{r}%
^{\prime }|}\;\mathrm{d}^{3}r\,\mathrm{d}^{3}r^{\prime },  \label{metric} \\
G^{MN}X_{N}^{\lambda } &=&\lambda X_{M};\ F^{\lambda }(\bm{r}%
)=X_{M}^{\lambda }F^{M}(\bm{r}).  \notag
\end{eqnarray}
\end{itemize}

We use the Coulomb metric, the favorable properties of which are well known
in quantum chemistry \cite{Pedersen-etal:2009}. Although the procedure is
carried out separately for each pair of atoms at a time, the overall result
is a basis in the space of all products. Our procedure provides us with (i)
a set of dominant functions $\{F^{\lambda }(\bm{r})\}$ and (ii) their
relation with the original products 
\begin{equation}
f^{a}(\bm{r})f^{b}(\bm{r})\sim \sum_{\lambda >\lambda _{\min }}V_{\lambda
}^{ab}F^{\lambda }(\bm{r}).  \label{ansatz-for-orbital-prod}
\end{equation}%
Here we ignore the products that have a Coulomb norm less than $%
\lambda_{\min }$. 
The finite support of the original atomic orbitals and the locality of our
construction are reflected in the sparse character of the ``vertex''
$V_{\lambda}^{ab}$.

Due to (i) the finite support of the original orbitals and (ii) the locality
of our procedure, identifying the dominant functions in a molecule of $N$
atoms takes asymptotically only $O(N)$ operations. Moreover, a very accurate
representation of orbital products as an expansion about intermediate points
of each pair is possible thanks to two powerful algorithms developed by
Talman \cite{Talman}. For the case of bilocal products, the results of our
procedure are illustrated in figure \ref{f:cc-prod}.

For reasons not yet entirely understood \cite{Larrue:2008}, the eigenvalues
of the metric (\ref{metric}) are asymptotically evenly spaced on a
logarithmic scale, in formal analogy with the $1/f$ noise that occurs in
electronics \cite{Brown:2007}. As a welcome consequence, the residual
(asymptotic) error of our algorithm vanishes exponentially fast with the
number of dominant products retained.

\section{Computation of the Kohn--Sham density response in $O(N^{2}N_{%
\protect\omega })$ operations}

\label{s:response}

The usual LCAO method provides us with a tensor basis for the electron
operators $c_{a}(t)$, $c_{b}^{+}(t^{\prime })$ and for their Green's
function $\mathrm{i}G_{ab}(t-t^{\prime })=T\langle
c_{a}(t)c_{b}^{+}(t^{\prime })\rangle $. Similarly, our dominant products
provide a tensor basis for the density 
\begin{eqnarray*}
n(\bm{r},t) &=&\psi ^{+}(\bm{r},t)\psi (\bm{r},t)=\sum_{\lambda }n_{\lambda
}(t)F^{\lambda }(\bm{r}) \\
\text{with \ }n_{\lambda }(t) &=&\sum_{a,b}c_{a}^{+}(t)V_{\lambda
}^{ab}c_{b}(t).
\end{eqnarray*}%
Recalling that the response function coincides with the density--density
correlator \cite{Fetter-Walecka:1971,Extra_comment} 
\begin{equation*}
\chi _{0}(\bm{r},\bm{r}^{\prime },t-t^{\prime })=-\mathrm{i}\langle T\{n(%
\bm{r},t)n(\bm{r}^{\prime },t^{\prime })\}\rangle _{\mathrm{connected}},
\end{equation*}%
we obtain a representation of the response function 
\begin{equation*}
\chi _{0}(\bm{r},\bm{r}^{\prime },t-t^{\prime })\equiv \sum_{\mu }F^{\mu }(%
\bm{r})\chi _{\mu \nu }^{0}(t-t^{\prime })F^{\nu }(\bm{r})
\end{equation*}%
in terms of a time-dependent matrix $\chi _{\mu \nu }^{0}(t-t^{\prime })$ 
\begin{equation}
\chi _{\mu \nu }^{0}(t-t^{\prime })=-\mathrm{i}\langle T\{n_{\mu }(t)n_{\nu
}(t^{\prime })\}\rangle _{\mathrm{connected}}
=\mathrm{i}\sum_{a,b,c,d}V_{\mu }^{ba}G_{ac}(t-t^{\prime })V_{\nu
}^{cd}G_{db}(t^{\prime }-t).  \label{response-via-gf}
\end{equation}%
%
\begin{figure}[h]
\centerline{\includegraphics[width=8cm,clip]{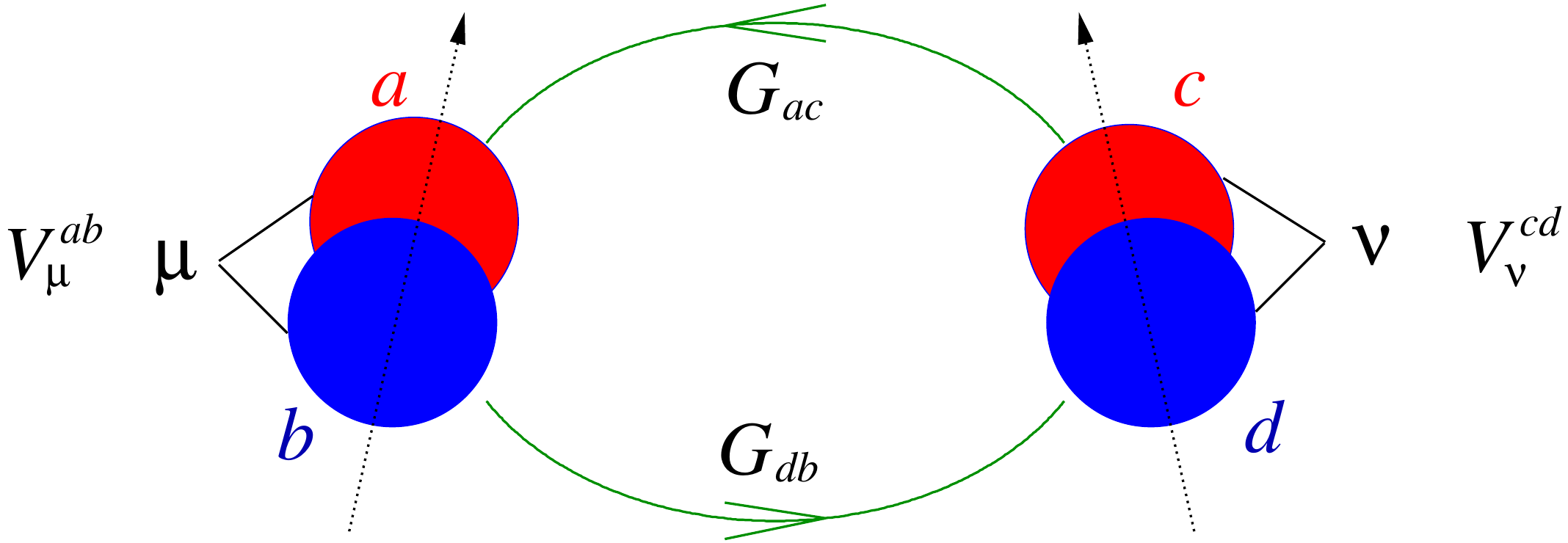}}
\caption{Particle--hole diagram for Kohn--Sham response in a basis of
dominant products. The vertex $V_{\protect\mu }^{ab}$ connects pairs of
orbitals $a,b$ to a dominant product $\protect\mu $. The propagators connect
the orbitals within the atom quadruplet. For a given pair of dominant
products $(\protect\mu ,\protect\nu )$, only the orbitals that belong to the
corresponding atom quadruplet must be summed over. Therefore, the total
computational effort scales as $O(N^{2}N_{\protect\omega })$.}
\label{f:particle--hole}
\end{figure}
From figure \ref{f:particle--hole} and from the locality of the vertex we
see that for $N$ $\gg 1$ atoms, the computation of $\chi _{\mu \nu
}^{0}(t-t^{\prime })$ takes $O(N^{2})$ operations. More precisely, the
number of operations is $O(N^{2}N_{\omega }\log (N_{\omega }))$ for $%
N_{\omega }$ frequencies where the $N_{\omega }\log (N_{\omega })$ factor is
due to the use of fast Fourier techniques in evaluating equation~(\ref%
{response-via-gf}).

To achieve a well controlled calculation, we express $\chi _{\mu \nu
}^{0}(\omega )$ in terms of its spectral representation 
\begin{equation}
\chi _{\mu \nu }^{0}(\omega )=\int_{-\infty }^{\infty }\mathrm{d}\lambda 
\frac{a_{\mu \nu }(\lambda )}{\omega -\lambda +\mathrm{i}\varepsilon },
\label{response-via-sf}
\end{equation}%
where $a_{\mu \nu }(\lambda )$ is a spectral function associated with
response function $\chi _{\mu \nu }^{0}(\omega )$. We now indicate how
expression (\ref{response-via-sf}) can be derived, for instance, by
representing the Green's functions in equation (\ref{response-via-gf}) in
terms of its spectral functions. We first recall the standard expression 
\cite{Fetter-Walecka:1971,Harrison:1968} for Green's function in terms of
molecular orbitals $\psi _{E}(\bm{r})$ 
\begin{equation*}
G(\bm{r},\bm{r}^{\prime },t-t^{\prime })=\mathrm{i}\theta (t^{\prime
}-t)\sum_{E<0}\psi _{E}(\bm{r})\psi _{E}(\bm{r}^{\prime })\mathrm{e}^{-%
\mathrm{i}E(t-t^{\prime })}
-\mathrm{i}\theta (t-t^{\prime })\sum_{E>0}\psi _{E}(\bm{r})\psi _{E}(\bm{r}%
^{\prime })\mathrm{e}^{-\mathrm{i}E(t-t^{\prime })}.
\end{equation*}%
We then rewrite it using the LCAO expression for the molecular orbitals $%
\psi _{E}(\bm{r})\equiv \sum_{a}X_{a}^{E}f^{a}(\bm{r})$ and introducing
density matrices of particles
$\rho_{ab}^{+}(s)=\sum_{E>0}X_{a}^{E}X_{b}^{E}\delta (s-E)$ and holes
$\rho_{ab}^{-}(s)=\sum_{E<0}X_{a}^{E}X_{b}^{E}\delta (s-E)$

\begin{equation*}
G_{ab}(t-t^{\prime })=\mathrm{i}\theta (t-t^{\prime })\int ds\ \rho
_{ab}^{-}(s)\mathrm{e}^{-\mathrm{i}s(t^{\prime }-t)}
-\mathrm{i}\theta (t-t^{\prime })\int ds\ \rho _{ab}^{+}(s)\mathrm{e}^{-%
\mathrm{i}s(t-t^{\prime })}.
\end{equation*}%
Inserting the last expression into equation (\ref{response-via-gf}), we
identify the spectral function $a_{\mu \nu }(\lambda )$ in equation (\ref%
{response-via-sf}) 
\begin{equation}
a_{\mu \nu }(\lambda )=\sum_{a,b,c,d}\left[ V_{\mu }^{ba}\rho
_{ac}^{+}\otimes V_{\nu }^{cd}\overline{\rho }_{db}^{-}\right] (\lambda ).
\label{sf-via-dm}
\end{equation}%
Here the overlined density $\overline{\rho }$ refers to a reflection of the
argument $\overline{\rho }(x)=\rho (-x)$ and the notation $\otimes $
represents the convolution.

The convolutions in equation (\ref{sf-via-dm}) can be done in $O(N_{\omega
}\log (N_{\omega }))$ operations using the FFT technique, provided the
density matrices $\rho _{ab}^{\pm }(s)$ are discretised on an uniform grid.
The discretisation of stick-like density matrices is done with a simple
algorithm---dividing the spectral weight between adjacent grid points
according to the distance between an eigenenergy $E$ and the grid point \cite%
{DF:2009}. The integration over $\lambda $ in equation (\ref{response-via-sf}%
) can also be represented as a convolution and may be computed in a fast manner
using the FFT technique \cite{DF:2009}.

We checked our results against the exact but slow expression of the
Kohn--Sham density response, i.~e. 
\begin{equation*}
\chi _{0}^{\mathrm{exact}}(\bm{r},\bm{r}^{\prime},\omega)
=\sum_{E,F}(n_{F}-n_{E})\frac{\psi _{E}(\bm{r})\psi _{F}(\bm{r})\psi _{E}(%
\bm{r}^{\prime})\psi _{F}(\bm{r}^{\prime})}{\omega -(E-F)-\mathrm{i}%
\varepsilon (n_{E}-n_{F})}
\end{equation*}%
and its corresponding expression in our tensor basis (see figure \ref%
{f:chi0_11} for a comparison between exact and fast results). 
This summarizes our method of constructing the Kohn--Sham density response
function $\chi _{0}$ in $O(N^{2}N_{\omega }\log (N_{\omega }))$ operations.
Next we will test our construction on molecular spectra.

\begin{figure*}[t]
\centerline{\includegraphics[width=16cm,viewport=70 575 485
725,clip]{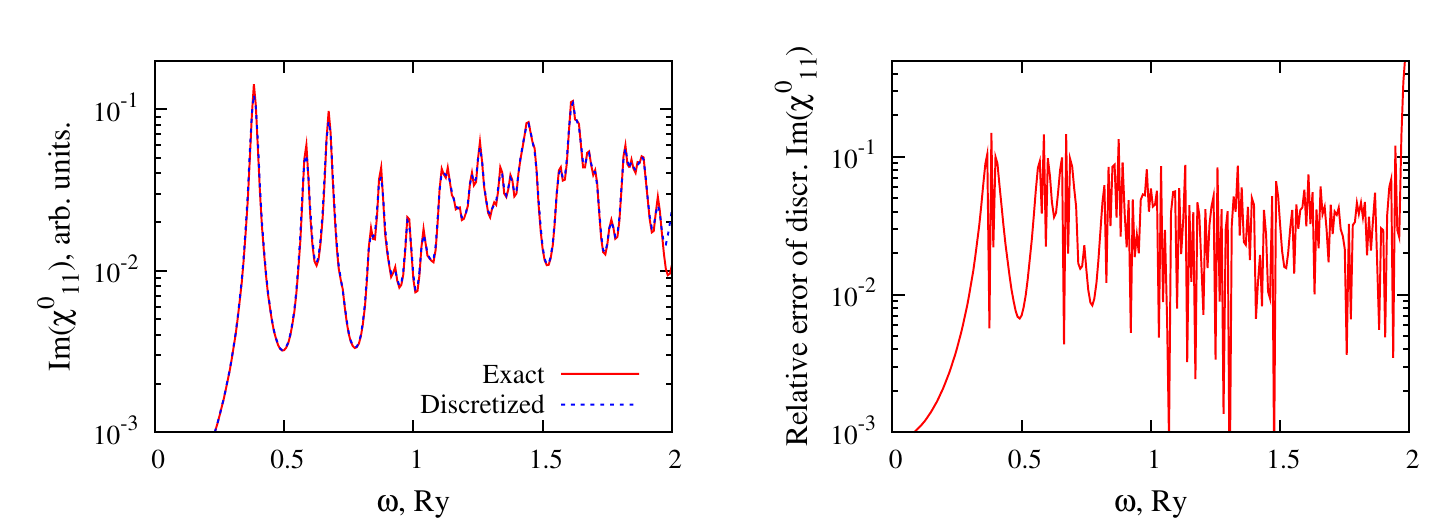}}
\caption{An element of the response function $\protect\chi^0_{\protect\mu%
\protect\nu}$ of benzene. The Kohn--Sham eigenstates were generated using
the SIESTA package \protect\cite{siesta} with default settings. The
discretised response function is computed in the frequency window $\protect%
\omega<\protect\omega_{\max}$, $\protect\omega_{\max}=2$ Rydberg with $N_{%
\protect\omega}=512$ data points. $\protect\varepsilon$ is chosen to be $1.5
\Delta\protect\omega$, where the discretisation spacing is $\Delta\protect%
\omega=2\protect\omega_{\max}/N_{\protect\omega}$.}
\label{f:chi0_11}
\end{figure*}

\section{Application of $\protect\chi_{0}$ to molecular spectra}

\label{s:tddft}

The most direct application of $\chi_{0}$ as constructed in the previous
section is the computation of excitation spectra of molecules. 

The excitation properties of molecules are determined by an interacting
response function $\chi $ that is related to the non interacting response
function $\chi _{0}$ \cite{Petersilka-etal:96} via a Dyson type equation

\begin{equation}
\chi =\frac{\delta n}{\delta V_{\mathrm{ext}}}=\frac{1}{1-f_{\mathrm{Hxc}%
}\chi _{0}}\chi _{0}.  \label{true-response}
\end{equation}%
The TDDFT kernel $f_{\mathrm{Hxc}}=\displaystyle\frac{\delta V_{\mathrm{Hxc}}}{\delta n}$
becomes a matrix in the basis of dominant products

\begin{equation}
f_{\mathrm{Hxc}}^{\mu \nu }= \iint 
\frac{F^{\mu }(\bm{r})F^{\nu }(\bm{r}^{\prime})}{|\bm{r}-\bm{r}^{\prime}|}\,%
\mathrm{d}^{3}r\mathrm{d}^{3}r^{\prime} \\
+\iint F^{\mu }(\bm{r})\frac{\delta V_{\mathrm{xc}}(\bm{r})}{\delta n(\bm{r}%
^{\prime})}F^{\nu }(\bm{r}^{\prime})\,\mathrm{d}^{3}r\mathrm{d}%
^{3}r^{\prime}.
\end{equation}%
In this work, we focus on the local density approximation (LDA) for the
exchange--correlation potential. In this case $f_{\mathrm{xc}}(\bm{r},\bm{r}%
^{\prime})=\frac{\mathrm{d}V_{\mathrm{xc}}(n)}{\mathrm{d}n}(\bm{r})\delta (%
\bm{r}-\bm{r}^{\prime})$ and its matrix elements are local in coordinate
space.

\subsection{Calculation of the interaction kernel $f_{\mathrm{Hxc}}$}

In the construction of dominant products, we distinguish between (i)
coincident and (ii) distinct or bilocal pairs of atoms. The former have full
rotational symmetry, while the latter have only axial symmetry with respect
to a line connecting their centers. Therefore the expansion of the bilocal
products is done in an appropriately rotated coordinate frame. Both local
and bilocal dominant products are expressed as a sum over angular--radial
functions in a suitable coordinate system

\begin{equation}
F^{\mu }(\bm{r})=\sum_{l}F_{l}^{\mu }(|\bm{r}-\bm{C}_{\mu }|)Y_{lm_{\mu }}(%
\bm{R}_{\mu }(\bm{r}-\bm{C}_{\mu })),
\end{equation}%
where the rotation $\bm{R}_{\mu }$ and the shift $\bm{C}_{\mu }$ are
determined by the atom pair. Using the theory of angular momentum, one can
reduce the Coulomb interaction $\iint \frac{F^{\mu }(\bm{r})F^{\nu }(\bm{r}%
^{\prime})}{|\bm{r}-\bm{r}^{\prime}|}\,\mathrm{d}^{3}r\mathrm{d}%
^{3}r^{\prime}$ to a sum over elementary two-center Coulomb integrals 
\begin{equation}
(\bm{C}|\bm{C}^{\prime})=\iint \frac{\mathrm{g}_{lm}(\bm{r}-\bm{C})\mathrm{g}%
_{l^{\prime}m^{\prime}}(\bm{r}^{\prime}-\bm{C}^{\prime})}{|\bm{r}-\bm{r}%
^{\prime}|}\,\mathrm{d}^{3}r\mathrm{d}^{3}r^{\prime}
\end{equation}%
between two functions of spherical symmetry $\mathrm{g}_{lm}(\bm{r})$. When
the orbitals overlap, the Coulomb integrals are computed in the momentum
space (where they become local) while, if they do not overlap, the Coulomb
interaction is calculated exactly in terms of moments. For converting
between coordinate and momentum space, we use Talman's fast Bessel transform 
\cite{Talman}.

Unlike the Hartree kernel, the LDA exchange--correla\-tion kernel is local
in coordinate space and a numerical integration is an appropriate procedure
for it. The integrand $F^{\mu }(\bm{r})f_{\mathrm{xc}}(\bm{r})F^{\nu }(\bm{r}%
)$ is of non trivial support due to the lens-shaped support of the dominant
products. However, the non spherical part of this volume is small for
neighboring atoms and we used spherical coordinates centered about a
midpoint between the centers of the dominant products $F^{\mu }(\bm{r})$, $%
F^{\nu }(\bm{r})$. The integration is done with a Gauss--Legendre method
along the radial coordinate and with a Lebedev quadrature \cite%
{Lebedev-theory-and-program} over the solid angle. A moderate number of
integration points leads to a sufficient accuracy in the
exchange--correlation matrix elements.

The computational complexity of the Hartree kernel $f_{\mathrm{H}}^{\mu \nu
} $ and the exchange--correlation kernel $f_{\mathrm{xc}}^{\mu \nu }$ are $%
O(N^{2})$ and $O(N)$, respectively. In practice, the calculation of the
kernel $f_{\mathrm{Hxc}}$ is faster than the calculation of the non
interacting response $\chi_{0}$ by an order of magnitude.

\subsection{An iterative method for finding the interacting polarizability}

\label{s:iter} Expression (\ref{true-response}) for the interacting response 
$\chi _{\mu \nu }$ involves matrix multiplications and inversion and this
would appear to require $O(N^{3}N_{\omega })$ operations, more than the $%
O(N^{2}N_{\omega })$ complexity of the non interacting response $\chi _{\mu
\nu }^{0}$. Fortunately, the polarizability $P$ is an average quantity%
\footnote{%
In this equation and below, we focus on the calculation of a particular
component of the polarizability tensor and remove tensor indices from the
polarizability $P(\omega )$ and Cartesian vector indices from the dipole
momenta $d_{i}^{\mu }=\int F^{\mu }(\bm{r})\bm{r}_{i}\,\mathrm{d}r$.
However, a corresponding block-Lanczos algorithm exists and allows for a
faster simultaneous calculation of all components of the polarizability tensor \cite%
{DF:2009}.} 
\begin{equation}
P(\omega )=\sum_{\mu ,\nu }d^{\mu }\chi _{\mu \nu }(\omega )d^{\nu },
\end{equation}
that is easy to compute using an iterative method of the Krylov type \cite%
{DF:2009}. A biorthogonal Lanczos method \cite{Saad} allows us to find a
simple representation for the (non hermitian) matrix $A=1-f_{\mathrm{Hxc}%
}\chi ^{0}$

\begin{equation}
A=\sum_{n,m}|L^{n}\rangle t_{nm}\langle R^{m}|,
\end{equation}%
where the matrix $t_{nm}$ is tridiagonal. The set of vectors $|L^{n}\rangle $
and $\langle R^{m}|$ build up the identity in the Krylov space, $\langle
R^{m}|L^{n}\rangle =\delta ^{mn}$. Therefore, if we choose the starting
vectors $|L^{1}\rangle $ and $\langle R^{1}|$ in the direction of the dipole
moment $d$ and of $\chi_{0}d$, respectively 
\begin{equation}
|L^{1}\rangle =d;\ \langle R^{1}|=\chi ^{0}d,
\end{equation}%
then the interacting polarizability takes a particularly simple form 
\begin{equation}
P(\omega )=t_{11}^{-1}\langle d|\chi _{0}(\omega )|d\rangle .
\end{equation}%
Here, we multiply the Kohn--Sham (non interacting) polarizability with the
first element of the inverse matrix $t_{nm}$.

The computational complexity of the iterative method is $O(N^{2}N_{\mathrm{K}%
})$, where $N_{\mathrm{K}}$ is the dimension of the Krylov subspace. In our
calculation $N_{\mathrm{K}}$ is very small compared to the number of
dominant products and grows slowly as we increase the size of the system. In
conclusion, our method requires a total of $O(N^{2}N_{\omega }
\log(N_{\omega}))$ operations to compute the molecular polarizability $%
P(\omega )$.

\subsection{Memory requirements}

The Kohn--Sham response matrix requires $O(N^{2}N_{\omega })$ memory. For
instance, in the case of indigo dye, we need approximately 4\thinspace
GBytes, where we have exploited the diagonal symmetry of the response
matrix, using single precision complex numbers, and calculating for 256
frequencies. This is beyond the capacity of most contemporary desktop
computers. The storage of the response matrix on a hard disk is not an
option because the matrix is generated element by element for all
frequencies at once, but needed (in the iterative computation of
polarizability) as a frequency dependent matrix. For this reason, we have to
employ parallel machines if a large molecule, say 100 atoms, is to be
treated. Currently, we are implementing OpenMP and MPI parallelized versions
of our algorithm.

\subsection{Some illustrative results}

We illustrate our method on benzene, indigo and fullerene C$_{60}$. Benzene
serves as a test whether our basis represents the non interacting response
function correctly. For indigo and fullerene, we compare our results with
those of ADF \cite{Te-Velde-etal:2001} and Quantum Espresso \cite%
{QE-official-citation}, respectively.

\begin{figure}[h]
\hfil\includegraphics[width=7.5cm,viewport=80 60 480 350]{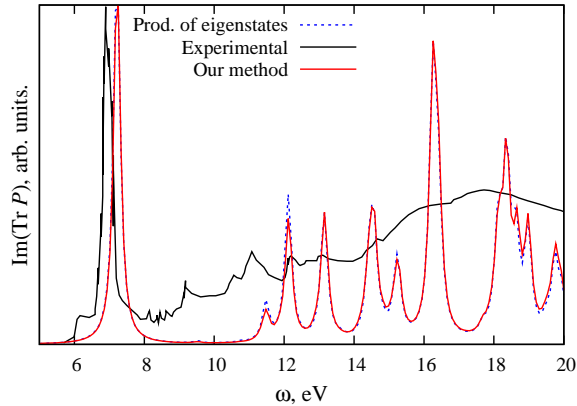}\hfil
\caption{The interacting polarizability of benzene computed with two
iterative methods compared with the experimental spectrum of benzene. The
calculations differ by their basis -- dominant products versus products of
eigenstates. The input of both calculations is generated in a SIESTA run.
Although different types of product basis are employed, the agreement
between the theoretical results is good. The experimental spectrum shows a
slightly red-shifted collective peak. See the text for a discussion
of possible reasons of this discrepancy.}
\label{f:benzene}
\end{figure}

In figure \ref{f:benzene} we compare two theoretical calculations with
experimental data \cite{Koch-Otto-1972} for benzene (C$_{6}$H$_{6}$). Both
calculations make use of the iterative method of section \ref{s:iter} but we
used (i) the basis of dominant products and (ii) the basis of products of
(Kohn--Sham) eigenstates in these calculations. In the basis of products of
eigenstates, the non interacting response $\chi _{0}$ is a diagonal matrix 
\cite{Martin-book} of size $N_{\mathrm{occ}}N_{\mathrm{virt}}\sim N^{2}$,
that is easy to compute. Therefore, the comparison (i) vs (ii) allows us to
assess whether there are enough dominant products for representing the
Kohn--Sham response function. The size of the set of products of
molecular orbitals grows as $N^{2}$ and the TDDFT kernel becomes a dense
matrix, leading to an overall complexity scaling $O(N^{3}N_{\omega })$ and
limiting the possibility of such a comparison to relatively small molecules
like benzene or naphthalene. The basis is chosen to contain 1238 dominant
products, which is considerable less than the 108*109/2=5886 original
products.

The good agreement of the two theoretical
calculations indicates the validity of our method as a whole and the
adequacy of the dominant product's basis for representing the Kohn--Sham
response function.

By comparison with experiment, the collective peak is slightly blue-shifted.
There are several reasons for this discrepancy. A first reason is the
limited applicability of the simple LDA functionals --- we use the
Perdew--Zunger LDA functional, which is the default functional in the SIESTA
package \cite{siesta}. A preliminary calculation with a GGA functional shows
better results. A second reason is the incompleteness of the LCAO basis set
--- we employed a double-zeta polarized (DZP) basis which is the default basis
in the SIESTA package. A third reason is our choice of the pseudo
potential. We have been using the pseudo potentials published on the SIESTA
Internet site \cite{SIESTA-site} (LDA potentials of Troullier-Martins type
adapted from ABINIT package \cite{ABINIT-site}). Surely one can perform a
fine tuning of the parameters, but this is beyond the scope of this paper.
Therefore, the remaining calculations presented below are done with the same
set of parameters.

\begin{figure}[h]
\hfil\includegraphics[width=7.5cm,viewport=80 60 480 350]{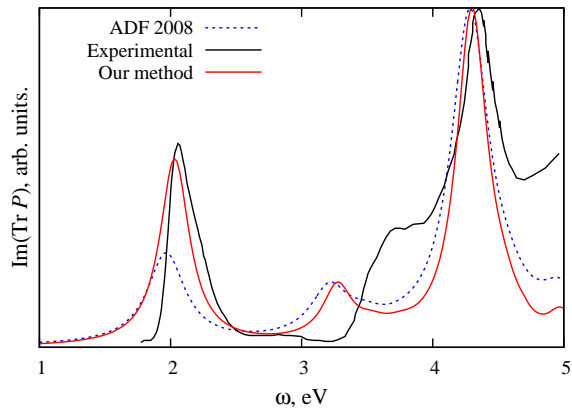}\hfil
\caption{Comparison of the absorption of indigo dye computed with the ADF
package versus our method. DFT calculations are done independently in ADF
and SIESTA. Nevertheless, theoretical spectra agree except for a factor of 2
in the height of the HOMO--LUMO peak. In the range between 3--4 eV
experiment and theory differ. The deviation might be caused by the presence
of solvent in experimental setup.}
\label{f:indigo}
\end{figure}

\begin{figure}[h]
\hfil\includegraphics[width=7.5cm,viewport=80 60 480 350]{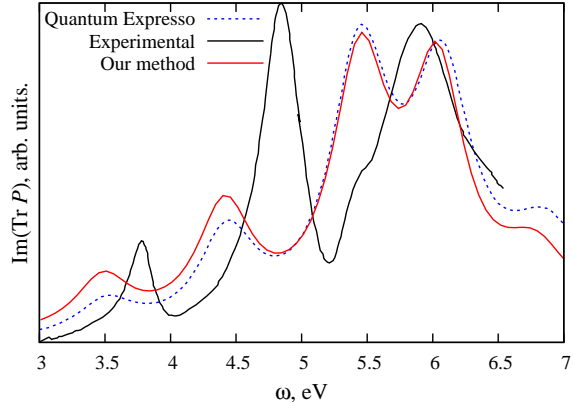}\hfil
\caption{Theory versus experiment for absorption of fullerene C$_{60}$.
Experimental data are from \protect\cite{Bauernschmitt-etal:1998} and we
compare our result with those of \protect\cite{Rocca-etal:2008}. The
theoretical results agree with each other and disagree with experiment. This
might indicate an inadequacy of the LDA exchange--correlation functional for
large molecules or be due to the presence of a solvent in the experimental
setup. Excitonic effects may be another reason for this discrepancy.}
\label{f:fullerene-c60}
\end{figure}

In figure \ref{f:indigo} we compare two theoretical calculations with
experimental data \cite{Brown:2008} for the indigo dye (C$_{16}$N$_{2}$O$%
_{2} $H$_{10}$). The first calculation is done within the ADF package \cite%
{Te-Velde-etal:2001} with parameters similar to SIESTA's default parameters,
while the second calculation is done with our method (details of the
calculation are collected above in the discussion of results for benzene).

Both calculations agree except for the strength of the HOMO--LUMO
transition. The experimental spectrum is in overall agreement with both
calculations: three resonances are seen, while the middle resonance is
probably disturbed by the presence of the solvent.

We kept about 2100 dominant products and 256 frequency points in this
calculation. The current implementation took 3.5 hours, if run on one
thread, on an Intel Xeon processor at 2.50\thinspace GHz.

In figure \ref{f:fullerene-c60} we compare theoretical and experimental
spectra for fullerene C$_{60}$. The calculation by Rocca \textit{et al} \cite%
{Rocca-etal:2008} agrees remarkably well with our calculation, while the
experimental results \cite{Bauernschmitt-etal:1998} are blue-shifted
compared with theory (details of the calculation are collected above in the
discussion of results for benzene).

The disagreement might be due to use of the LDA functional. A solvent usually
gives rise to a uniform red shift of experimental data \cite{Brown:2009}
which is not the case in this example. We believe that excitonic effects
cause the discrepancy in this large molecule.

We kept about 8700 dominant products and used 128 frequency points. The
current implementation requires 18.4 hours, if run on one thread, on an
Intel Xeon processor at 2.50\thinspace GHz.

The agreement between our calculations with those done by other authors and
with experiment validates our construction of a basis in the space of
products and our construction of the Kohn--Sham response function $\chi_{0}$.

\section{Hedin's GW self-energy in the product basis}

\label{s:gw}
It is known that particle--hole interactions in organic semiconductors
cannot be described adequately by TDDFT. Such systems, however, 
can be modelled by Hedin's GW approximation \cite{Tiago,Sottile-etal:2007}.

Let us show how to express Hedin's GW self-energy \cite{Hedin:1965} 
in terms of our concepts $\{F^{\lambda }(\bm{r}),V_{\lambda }^{ab}\}$.
Although the self-energy operator $\Sigma$ should, in principle,
be determined self-consistently using Hedin's set of
integral equations, even a ``one shot'' approximation to the self-energy has
been shown to improve the quasi particle energies compared to TDDFT
energies. The self energy is given by \cite{Hedin:1965,Rieger-etal:1999}

\begin{equation}
\Sigma (\bm{r},\bm{r}^{\prime },t-t^{\prime })=\mathrm{i}G_{0}(\bm{r},\bm{r}%
^{\prime },t-t^{\prime })W(\bm{r},\bm{r}^{\prime },t-t^{\prime }),
\end{equation}%
where $W=\frac{1}{1-f_{\mathrm{H}}\chi _{0}}f_{\mathrm{H}}$ is a
RPA-screened Coulomb interaction that makes use of the non interacting KS
response $\chi _{0}$ and $G_{0}$ is the non interacting Green's function. To
see the form this equation takes in our basis of dominant functions, we
define a self energy matrix and expand the Greens function $G_{0}$

\begin{eqnarray*}
\Sigma ^{ab}(t-t^{\prime}) &=&\int \mathrm{d}^3r \mathrm{d}^3r^{\prime}f^{a}(%
\bm{r}) \Sigma(\bm{r},\bm{r}^{\prime},t-t^{\prime})f^{b}(\bm{r}^{\prime}) \\
G_{0}(\bm{r},\bm{r}^{\prime},t-t^{\prime})
&=&\sum_{a,b}G^0_{ab}(t-t^{\prime})f^{a}(\bm{r})f^{b}(\bm{r}^{\prime}).
\end{eqnarray*}%
Inserting $\Sigma =\mathrm{i}G_{0}W$ and using the representation of 
products (\ref{ansatz-for-orbital-prod}) we find the following tensor form
of the GW approximation
\begin{equation*}
\Sigma ^{ab}(t-t^{\prime}) \\
=\mathrm{i}\sum_{\mu,\nu,a^{\prime},b^{\prime}}V_{\mu }^{aa^{\prime}}V_{\nu
}^{bb^{\prime}} G^0_{a^{\prime}b^{\prime}}(t-t^{\prime})W^{\mu \nu
}(t-t^{\prime}),
\end{equation*}
\begin{eqnarray*}
W^{\mu \nu }(t-t^{\prime}) &=&\int \mathrm{d}^3r \mathrm{d}^3r^{\prime}\,
F^{\mu}(\bm{r})W(\bm{r},\bm{r}^{\prime},t-t^{\prime})F^{\nu }(\bm{r}%
^{\prime}).
\end{eqnarray*}

\begin{figure}[h]
\centerline{\includegraphics[width=7.5cm,clip]{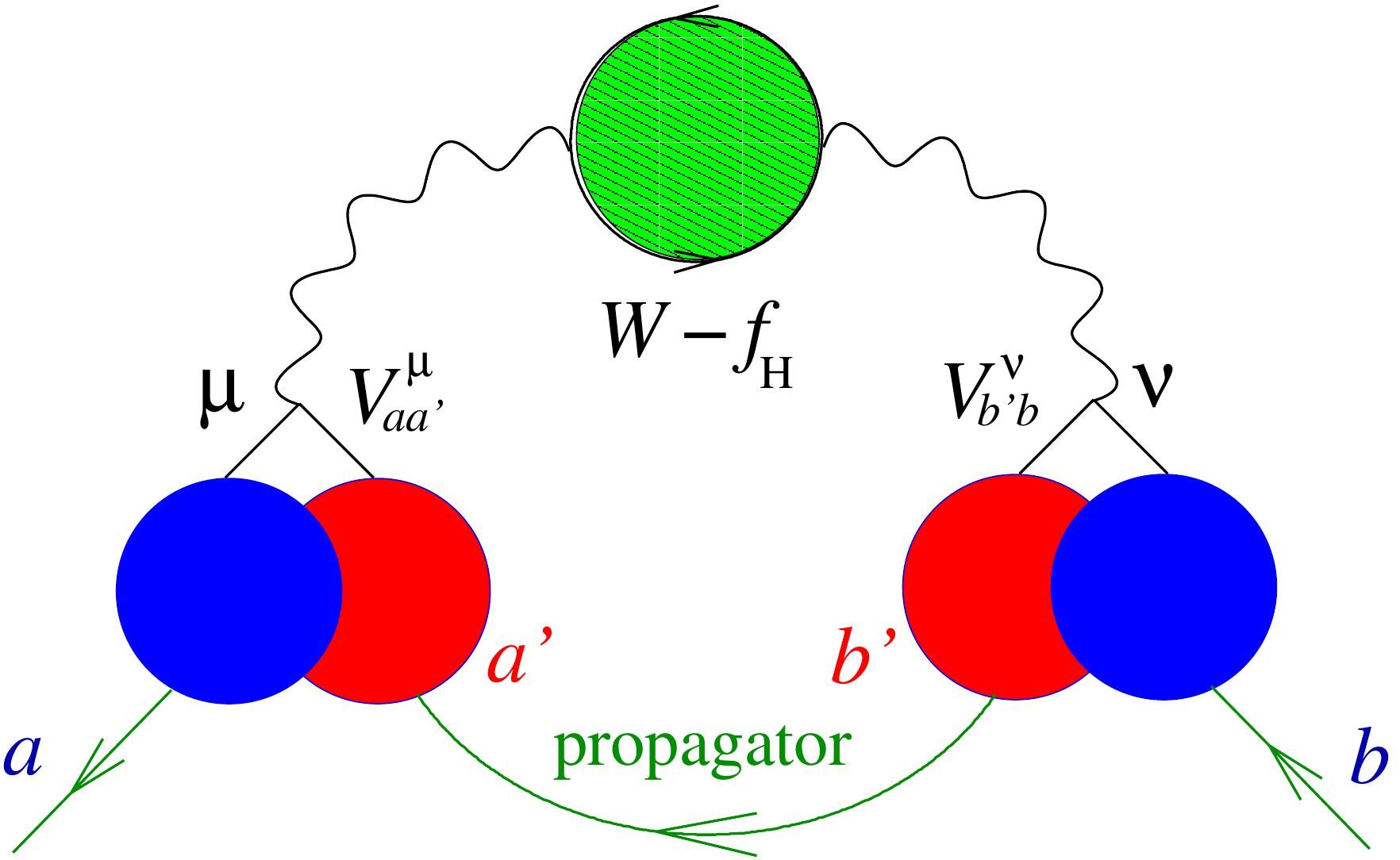}}
\caption{Diagram for the self-energy $\Sigma$ in a ``one shot'' GW approximation.
The vertices $V^{aa^{\prime}}_{\protect\mu}$ and $V^{b^{\prime}b}_{\protect%
\nu}$ are sparse -- only the orbitals and products belonging to a given
quadruplet of atoms contribute to the trace. Therefore, the computational
effort for self energy $\Sigma$ scales as $O(N^2 N_{\protect\omega})$ once
the screened Coulomb interaction is known.}
\label{f:self-energy-prod-basis}
\end{figure}

The direct computation of the screened Coulomb interaction $W^{\mu\nu}$
requires \textit{a priori} $O(N^{3}N_{\omega })$ operations while the rest
of the calculation of $\Sigma ^{ab}$ scales as $O(N^{2}N_{\omega })$. This
is due to the locality of the vertex $V_{\mu}^{ab}$ (see the diagrammatic
representation of the self-energy in figure \ref{f:self-energy-prod-basis}).
The situation here is similar as in the calculation of $\chi_{\mu \nu }^{0}$.

\bigskip

\section{Conclusions}

In this paper we reviewed our extension of the LCAO method to densities and
excited states and we gave first applications of this method. One
application is a convenient construction of the non interacting Kohn--Sham
response in $O(N^{2}N_{\omega })$ operations. Another application is the use
of this response function to compute electronic excitation spectra within
TDDFT linear response, again in $O(N^{2}N_{\omega })$ operations. We
illustrated our method of computing spectra using benzene, indigo and
fullerene and we also confirmed the $O(N^{2}N_{\omega })$ complexity scaling
of our method. A drawback of our construction of $\chi _{0}$ is its high
memory requirement--- one could use a MPI parallelization to address this
problem. We also mentioned that our method is suitable for application to
the GW approximation which we cast in an appropriate tensor form.

Because our method provides (i) a simple basis for the electronic density
and (ii) the Kohn--Sham response function we believe it will be useful in treating
excitonic effects 
in molecular physics.

\bigskip

\section*{Acknowledgement}

It is our pleasure to thank James Talman (University of Western Ontario,
London) for contributing two crucial algorithms to this project, for making
unpublished computer codes of these algorithms available to us, for many
fruitful discussions and for useful correspondence.

D.F. is grateful to Peter Fulde for extensive and continued support and for
inspiring visits at MPIPKS, Dresden that provided perspective for the
present work. Part of the collaboration with James Talman was done in the
pleasant environment of MPIPKS. D.F. acknowledges the kind hospitality
extended to him by Gianaurelio Cuniberti at the Nanophysics Center of
Dresden.

Both of us are indebted to Daniel S\'a{}nchez-Portal (DIPC, Donostia) for
strong support of this project and for advice and help on the SIESTA code.
We also thank Andrei Postnikov (Paul Verlaine University, Metz) for useful
advice.

Ross Brown (IPREM, Pau) helped with experimental
data on indigo dye and discussions. We acknowledge useful advice by Isabelle
Baraille (IPREM, Pau), Nguyen Ky and Pierre Gay (DRIMM, Bordeaux), and Alain
Marbeuf (CPMOH, Bordeaux).

We thank Mark~E.~Casida and Bhaarathi Natarajan (Joseph Fourier University,
Grenoble) and their colleagues at Centro de Investigacion, Mexico for
letting us use their deMon2k code and for much help with it. Stan van
Gisbergen's (Vrije Universiteit, Amsterdam) provided us with a trial license
of ADF and gave us useful comments on the algorithms implemented in ADF.

This work was financed by the French ANR project ``NOSSI'' (Nouveaux Outils
pour la Simulation de Solides et Interfaces). Financial support and
encouragement by ``Groupement de Recherche GdR-DFT++'' is gratefully
acknowledged. 


\begin{thebibliography}{99}
\bibitem{Extra_comment} For convenience, we use time-ordered, rather than
causal correlators.
\end{thebibliography}
\end{document}